\documentclass[a4paper,11pt]{article}
\pdfoutput=1
\usepackage{jcappub} 

\usepackage[T1]{fontenc} 

\usepackage{graphicx}
\usepackage{amssymb}

\title{\boldmath Magnetogenesis in Matter - Ekpyrotic Bouncing Cosmology}

\author {Ratna Koley,} 
\author {Sidhartha Samtani}

\affiliation {Department of Physics, Presidency University, 86/1 College Street, Kolkata, India, 700073}

\emailAdd{ratna.physics@presiuniv.ac.in}
\emailAdd{samtanisidhartha@gmail.com}

\abstract{${}$ 
In the recent past there have been many attempts to associate the
 generation of primordial magnetic seed fields with the inflationary era,
  but with limited success. We thus take a different approach by using a model for nonsingular bouncing cosmology. 
  A coupling of the electromagnetic Lagrangian  $F_{\mu\nu}F^{\mu\nu}$ with a non background scalar field 
  has been considered for the breaking of conformal invariance. 
  We have shown that non singular bouncing cosmology supports magnetogenesis evading the long standing 
  back reaction and strong coupling problems which have plagued 
      inflationary magnetogenesis. In this model, we have achieved a scale invariant power spectrum for the  
      parameter range compatible with observed CMB anisotropies. The desired strength of the magnetic field 
      has also been obtained that goes in accordance with present observations. It is also important to note that no BKL instability arises within this parameter range. The energy scales for different stages 
      of evolution of the bouncing model are so chosen that they solve the problems of Standard Big Bang cosmology as well.  
  }

  \arxivnumber{1612.08556}

\begin{document}
\maketitle
\flushbottom
\section{Introduction}\label{intro}
Magnetic fields have been observed on all scales probed so far, from planets and stars to the large scale magnetic fields in galaxies and 
galaxy clusters. The strength of the magnetic field detected in nearby galaxies ranges from a few to tens of micro Gauss
lined up in a coherence scale of the order of ten kpc \cite{Beck:2000dc} whereas the same in the galaxy clusters have been found to be of 
few micro Gauss coherent in kpc scales \cite{Vogt:2005xf}. A lower bound, $B_0 \gtrsim 3 \times 10^{-16}$~Gauss, on the strength of intergalactic 
magnetic fields has been imposed to explain the non observation of GeV gamma-ray emission from electromagnetic cascade 
initiated by TeV blazars in the intergalactic medium \cite{2010Sci...328...73N,Taylor:2011bn}.  Magnetic fields in galaxies and clusters are 
produced via the amplification of {\it{seed}} magnetic fields of unknown nature.  The only potential opportunity for searching the initial {\it{seed}}
field is to look for places in the Universe where these fields might exist in their original form, not distorted by the complicated plasma and
magneto hydrodynamic processes.  Intergalactic medium, more precisely, the voids of large-scale structure is the only place where such 
primordial magnetic fields might reside. Discovery of intergalactic magnetic fields has prompted many researchers to look for the origin 
of {\it{seed}} fields in early Universe, during epochs preceding the structure formation. On observational ground the main 
constraint on primordial {\it{seed}} fields comes from the variety of additional signals they add to the cosmic microwave background (CMB) radiation. 
Recent observational data from Planck 2015 put an upper bound of $\sim  2 \times 10^{-9}$~Gauss, on the  primordial magnetic fields 
\cite{Ade:2015cva, Paoletti:2010rx}. Researchers are trying for long to understand the primordial magnetic fields but there does not exist a completely 
satisfactory model till today. 

The idea of introducing primordial {\it{seed}} magnetic fields has been motivated by the successful generation of matter structure formation by quantum 
fluctuations of the scalar inflaton.  It is thus found to be interesting to explore whether large scale magnetic fields have the similar origin! There are several 
issues both theoretical and observational that one needs to address in one go to establish such a possibility. For example, in the minimal set up it turns 
out that the Maxwell electromagnetism is conformally invariant in a Friedmann-Robertson-Walker spacetime. As a result the vector field fluctuation 
modes do not feel the fast rate of expansion of the Universe during the inflation. So this mechanism does not lead to the generation of large scale 
magnetic fields. To resolve this issue Turner and Widrow \cite{PhysRevD.37.2743} considered a coupling of the electromagnetic Lagrangian 
$F_{\mu\nu}F^{\mu\nu}$ with gravitational field through Ricci scalar. But this model did not meet the expectation of the observable magnetic fields.
Later, it was suggested in \cite{Ratra:1991bn} to consider a kinetic coupling with the electromagnetic Lagrangian.  Recently, a lot of work have
been done to explain the generation of magnetic fields during the inflationary epoch. However, most of these models either predict magnetic 
fields which are extremely weak to match with observations or suffer from issues like backreation (generated electromagnetic energy 
density becomes too high to spoil the inflationary background) and strong coupling 
 (that the coupling with produced electromagnetic field may become so large that one is then in a strongly coupled theory at the 
 beginning of inflation where such a theory is not trustable) problems. Comprehensive 
review on inflationary magnetogenesis can be found in \cite{Subramanian:2015lua,Durrer:2013pga}. Recent works such as 
\cite{Ferreira:2013sqa, Ferreira:2014hma, Fujita:2015iga} have addressed the problems of inflationary magnetogenesis with some 
success.

It is well known that the inflationary scenario is the most accepted current paradigm of early universe cosmology. Although inflation solves 
the major problems of Standard Big Bang cosmology and gives rise to a causal theory of structure formation, it faces a number of conceptual 
challenges \cite{Brandenberger:1999sw}. For example, in the models 
where inflation is governed by the dynamics of a scalar field coupled to Einstein 
gravity  it can be shown, applying the Hawking-Penrose singularity theorems \cite{Hawking:1969sw} and their extensions 
\cite{Borde:1993xh}, that there is necessarily a singularity before the onset of inflation. This in turn tells that the inflationary scenario 
cannot yield the complete history of the very early universe. The other pathology arises from the scale of quantum fluctuations 
in the inflaton. In the inflationary cosmology if the period of inflation lasts slightly longer than the minimal duration, the scales of
cosmological interest today must have to originate in the sub Planckian regime where both General Relativity and quantum field 
theory break down \cite{Martin:2000xs, Banerjee:2016arl}. Bouncing cosmological scenario provides an 
alternative to inflationary scenario, bypassing the singularity problem. Also the bouncing cosmological models naturally avoid 
the trans Planckian problem because the scales of fluctuation we observe today originate at a  very high value compared to the Planck 
scale.  The model introduces a contracting phase for the Universe 
which undergoes a bounce, where the scale factor reaches a finite minima and then enters the standard expansion phase 
\cite{Cai:2014bea,Brandenberger:2009jq,Novello:2008ra,Brandenberger:2016vhg}. 
Irrespective of the above mentioned pathologies one can find it interesting to explore the bouncing cosmology. Because 
the major success of inflation in seeding large scale structure formation can also be achieved in this scenario. Simply observing gravitational 
waves on cosmological scales from B-mode polarization of CMB will not uniquely determine that inflation had happened 
\cite{Brandenberger:2011eq}. Rather it will be interesting to explore the alternative ideas and look for distinctive features that they give 
to come to a consensus about the actual model of early universe. It is also interesting to investigate in which models of cosmology the bouncing scenario can be accommodated \cite{0264-9381-23-23-001}.


In this article we address magnetogenesis in bouncing cosmology. A coupling of electromagnetic field with a non background
scalar is generally considered in this scenario. First such an attempt to generate scale invariant magnetic fields with sufficient strength was reported in 
\cite{Membiela:2013cea}. It was found that successful magnetogenesis can be achieved but at the cost of the possibility of having 
instabilities arising in the final expanding phase for some classes of steep Ekpyrotic potentials. An attempt has been made in 
\cite{Sriramkumar:2015yza} 
where no analytical solution for the mode functions has been achieved and numerically it has been pointed out that the back reaction problem 
may become severe in some sub classes where one can get analytical solutions \cite{Chowdhury:2016aet}. Recently the process of 
magnetogenesis in the context of nonsingular bounce cosmology has been explored in \cite{Qian:2016lbf}. Magnetogenesis took 
place in the Ekpyrotic phase and an instantaneous bounce and succeeding fast roll have been considered in this model. It has been shown that large 
primordial magnetic fields can be generated during the contraction phase without encountering strong coupling and back reaction issues. 
In the current work we consider the model of single field non singular bouncing cosmology developed in \cite{Cai:2012va}. 
In previous attempts authors
have chosen to study the evolution of magnetic fields only during the Ekpyrotic phase. Here we present a more generalized and complete 
picture by considering the coupling of the magnetic field starting from the matter dominated era (in the early contracting phase) and 
evolving through the Ekpyrotic phase.  


The paper is organized in the following way. We start with a motivation for the type of bouncing model we consider for the study of 
magnetogenesis followed by a quick discussion on generation of vector fluctuation modes and coupling function. We shall also discuss 
about the equation of motion. The next section is devoted to report different phases of bouncing cosmology and  the subsequent evolution 
of magnetic field spectra. In section 4, we calculate the electric and magnetic power spectra. After that we examine whether 
the void magnetic fields observed today can be generated in the non singular bouncing cosmology model. In section 6, we elaborate on the back reaction issue 
in this scenario. Finally we conclude with a discussion on the results obtained and a brief outlook on open issues.

\section{Model for Magnetogenesis}
\subsection{Non singular bounce model}

Let us briefly describe the model that we have considered. 
We follow the model of non singular bouncing cosmology 
with a single scalar field $\phi$ with a non-standard kinetic term and a non-trivial potential $V$ \cite{Cai:2012va, Cai:2013vm} 
and a pressure less matter fluid. 
The action of the model is given by,
\begin{equation}\label{eqn:2.1}
S=\int d^4x\sqrt{-g}\left[\frac{1}{2}R+\mathcal{L}_{bounce}-\frac{1}{4}I^2(a(\tau))F_{\mu\nu}F^{\mu\nu} \right]
\end{equation} 
where $R$ is the Ricci scalar, 
$F_{\mu\nu}=\partial_\mu A_\nu - \partial_\nu A_\mu$, the standard electromagnetic field tensor and 
$\mathcal{L}_{bounce}$ is the Lagrangian density associated with the non singular bouncing scenario and $I$ is an 
arbitrary function of the non background scalar field. Here we will consider $I(\tau)$ in such way that it is a power law function 
of the scale factor, $a(\tau)$. Let us now briefly describe this model of bouncing cosmology. We have already mentioned that 
conformal invariance should be broken in order to generate the magnetic field and here $I$ is the function that is playing the said role. 
Now the effective field theory Lagrangian, $\mathcal{L}_{bounce}$,  is given by
\begin{equation}\label{eqn:2.2}
\mathcal{L}_{bounce}=-K(\phi,X)-G(\phi,X)\Box \phi
\end{equation}
where $X$ is defined as the regular scalar kinetic term,  $X=\frac{1}{2}g^{\mu\nu}(\partial_\mu \phi)(\partial_\nu \phi)$, 
$K$ and $G$ are dimensionless functions of background scalar field $\phi$. 
\begin{align}
K(\phi,X)&=(1-g(\phi)X+\beta X^2-V(\phi)\label{eqn:2.3})\\
G(X)&=\Lambda X\label{eqn:2.4}
\end{align}
where $\beta$ and $\Lambda$ are parameters of the model. We also have the expressions of $V(\phi)$ and $g(\phi)$ which are given as,
\begin{align}
V(\phi)&=-\frac{2V_0}{e^{-\sqrt{\frac{2}{p}}\phi}+e^{b_V\sqrt{\frac{2}{p}}\phi}}\label{eqn:2.5}\\
g(\phi)&=\frac{2g_0}{e^{-\sqrt{\frac{2}{q}}\phi}+e^{b_V\sqrt{\frac{2}{q}}\phi}}\label{eqn:2.6}
\end{align}
where $V_0$ is positive and has the dimension of fourth power of $\mbox{mass}$. The other parameters  $b_V$, $p$, $q$ and $g_0$ carry positive 
values constrained as  $p<1$, $q<1$ and $g_0>1$ \cite{Cai:2012va, Cai:2013vm}. These constraints arise from the study of anisotropy 
and limit from CMB data on the  primordial sources of fluctuations.   


\subsection{Equation of motion, mode functions and coupling function}

The equation of motion governing the gauge field dynamics is obtained from the variation of the above Lagrangian 
in Eq. (\ref{eqn:2.1}) with respect to field variable $A_\mu$. 
\begin{equation}\label{eqn:2.8}
\frac{1}{\sqrt{-g}}\partial_\mu[\sqrt{-g} I^2(\tau)F^{\mu\nu}]=0
\end{equation}
We have considered a (3+1) dimensional spatially flat FRW metric described by the following line element.
\begin{equation}\label{eqn:2.7}
ds^2=a^2(\tau)(-d\tau^2+d\vec{x}^2)
\end{equation}
where $\tau$ is the conformal time and $a(\tau)$ is the scale factor as also mentioned earlier. We now propose a gauge choice 
and expand the gauge field $A_\mu$ in Fourier space for quantisation. Let us make the following choice wherein 
$A_0=0 ~\mbox{and} ~\partial_iA^i=0.$
In this gauge the spatial components of the vector field, $A^{i}$ or $\vec{A}$, can be expanded in Fourier space in the following way 
\begin{equation}\label{eqn:2.10}
\vec{A}(\tau,\vec{x})=\sum_{\lambda=1,2} \int \frac{d^3k}{(2 \pi)^{\frac{3}{2}}} 
\left[\vec{\epsilon}_\lambda(\vec{k})\hat{a}_\lambda(\vec{k})A_{k}(\tau) e^{i\vec{k}.\vec{x}}+\text{h.c.} \right]
\end{equation}
where the quantities $\vec{\epsilon}_\lambda (\vec{k})$ are the normalized transverse polarization vectors. 
By $\lambda$ the different polarization states have been represented here. We have defined the commutation 
relation of $\hat{a}_\lambda(\vec{k})$ 
and $\hat{a}^{\dagger}_{\lambda}(\vec{k})$ operators as,
\begin{equation}\label{eqn:2.11}
[a_{\lambda}(\vec{k}),a^{\dagger}_{\lambda'}(\vec{k'})] = \delta_{\lambda\lambda'}\delta(\vec{k}-\vec{k'})
\end{equation}
Thus the equation of motion for the spatial Fourier components reduces to
\begin{equation}\label{eqn:2.12}
A''_k+2 \frac{I'}{I}A'_k+k^2A_k=0
\end{equation}
where prime denotes derivative w.r.t. conformal time. One can further simplify the above equation of motion by the following redefinition, $\mathcal{A}_k =I A_k$. Note that with 
this the following equation looks like the Mukhanov-Sasaki equation in case of scalar mode. 
\begin{equation}\label{eqn:2.13}
\mathcal{A}''_k+\left(k^2-\frac{I''}{I} \right)\mathcal{A}_k=0
\end{equation}
It is apparent from the above equation that the coupling term acts as an effective mass term and is responsible for breaking the conformal
invariance. This is why when one wants to get back standard electromagnetism this coupling must be reduced to one. Solving the above 
eq. (\ref{eqn:2.13}) we will explicitly read the evolution of magnetic field which depends on the coupling function and different phases 
of background evolution. In the following section we will discuss how the mode functions evolve through different phases of background
evolution with appropriate matching conditions.

We will explicitly consider  the form of $I(\tau)$ as 
\begin{equation}\label{eqn:2.14}
I(\tau)=\Big[\frac{a(\tau)}{a_{B-}}\Big]^n
\end{equation}
where $a_{B-}$ is the scale factor at the end of the Ekpyrotic era. The coupling starts from the matter dominated phase and 
we can see from \eqref{eqn:2.14} that at the end of the Ekpyrotic phase the coupling goes to unity and we recover Maxwell electrodynamics. 
One can also observe that the coupling term is greater than one if we consider $n>0$, since the universe is contracting the scale factor 
in the matter dominated phase is greater than the following Ekpyrotic phase which makes the coupling term greater than unity for positive $n$. 
This quite naturally avoids the well known strong coupling problem (the coupling must be greater than unity to avoid strong coupling problem).  
Now we will consider the evolution of the magnetic fields through matter dominated contraction phase to Ekpyrotic phase. However, we will treat 
the bouncing phase and fast role expansion as instantaneous since these phases do not play a significant role \cite{Quintin:2015rta}.

\section{Different phases of background evolution and junction conditions}

The basic framework of non singular bouncing cosmology may be based on different theoretical considerations. For example, it 
may be developed on modified sector of gravity or in the framework of Einstein gravity by considering a matter sector that violates 
Null Energy Condition giving rise to bouncing phase or one can also get bounce with a non flat background spacetime 
\cite{PhysRevD.68.103517, 0264-9381-23-23-001}. 
In this work we exercise the second option mentioned above. However, an important issue in this case is the Belinsky-Khalatnikov-Lifshitz (BKL) instability.
This instability appears when the back reaction corresponding to the anisotropies overshoots the regular background densities unless a very finely tuned 
initial isotropic condition is considered. However there is another way out to tackle this issue by considering an Ekpyrotic phase before the bounce. The model 
we work with produces non singular bouncing cosmology from regular matter contraction via an era of Ekpyrotic phase driven by a scalar field with Hordensky 
type non standard kinetic term and a negative potential as mentioned in previous section. 
Let us now discuss how the magnetic field evolves with background evolution by applying appropriate
boundary conditions.

\subsection{Mode function in vacuum phase}
At very early times when conformal time, $\tau\rightarrow-\infty$ the gauge field does not feel any effect of the curvature of spacetime and 
we can  consider that the spacetime is effectively like the Minkowskian and choose the Bunch Davies vacuum solution to represent 
the gauge field $\mathcal{A}_k$ as 
\begin{equation}\label{eqn:3.1}
\mathcal{A}_k (\tau) = \frac{e^{-ik\tau}}{\sqrt{2k}}
\end{equation}
where $k = 2\pi/\lambda$ and it represents the wave number corresponding to a fluctuation mode of wavelength $\lambda$. 
\subsection{In initial matter contraction phase}
In the non singular bouncing cosmology model mentioned above the scale factor during the initial matter 
contraction phase (dominated by a pressureless matter fluid) has been considered to be  
\begin{equation}\label{eqn:3.2}
a(\tau) = a_E \Big(\frac{\tau-\tilde{\tau}_E}{\tau_E-\tilde{\tau}_E}\Big)^2
\end{equation}
where
$\tau_E-\tilde{\tau_E}=\frac{2}{a_EH_E}$ and
$a_E\equiv a(\tau_E)$ is the scale factor at the end of the matter dominated phase ($\tau_E$). Here $H_{E}$ is the Hubble parameter at $t_{E}$, where $t_{E}$ is the cosmic time when the matter dominated phase ends. With this choice the equation of motion given by 
eq. \eqref{eqn:2.13} produces the following differential equation second order in conformal time.
\begin{equation}\label{eqn:3.3}
\mathcal{A}_k'' + \Big(k^2-\frac{2n(2n-1)}{(\tau -\tilde{\tau}_E)^2}\Big)\mathcal{A}_k=0
\end{equation}
General solution to this equation can be written as a linear combination of two independent solutions as follows. 
\begin{equation}\label{eqn:3.4}
\mathcal{A}_k(\tau)=\sqrt{-(\tau -\tilde{\tau}_E)}\Big[C_1H^{(1)}_{\frac{4n-1}{2}}(-k(\tau-\tilde{\tau}_E))+C_2H^{(2)}_\frac{4n-1}{2}
(-k(\tau-\tilde{\tau}_E)) \Big]
\end{equation}
where $H_\nu^{(1)}$ and $H_\nu^{(2)}$ are Hankel functions of the first and second kind respectively and $C_1$ and $C_2$ are two constant 
coefficients which will be found from junction conditions. In the limit $\tau \rightarrow -\infty$ the above solution must agree with 
that in vacuum phase in Eq.~\eqref{eqn:3.1}. Hence the above expression reduces to the following form after applying appropriate 
boundary condition. 
\begin{equation}\label{eqn:3.5}
\mathcal{A}_k(\tau)=\sqrt{\frac{-\pi(\tau-\tilde{\tau}_{E})}{4}}H^{(1)}_{\frac{4n-1}{2}}(-k(\tau -\tilde{\tau}_{E}))
\end{equation}
Note that the negative sign inside the square root does not create any tension because the conformal time itself 
is a negative quantity in this regime. We now move on to the Ekpyrotic phase.
\subsection{Gauge field in Ekpyrotic phase}
The expansion scale factor during the Ekpyrotic phase is given by a generalized power law of conformal time. 
\begin{equation}\label{eqn:3.6}
a(\tau)=a_{B-}\Big(\frac{\tau-\tilde{\tau}_{B-}}{\tau_{B-}-\tilde{\tau}_{B-}}\Big)^{\frac{p}{1-p}}
\end{equation}
where $\tau_{B-}-\tilde{\tau}_{B-}=\Big(\frac{p}{(1-p)a_{B-}H_{B-}}\Big)$ and $a_{B-} \equiv a(\tau_{B-})$ i.e. 
the scale factor at the end Ekpyrotic phase (i.e. ${\tau}_{B-}$). Here $H_{B-}$ is the Hubble parameter at $t_{B-}$, where $t_{B-}$ is the cosmic time when Ekpyrotic phase ends.
The constraint on the parameter $p$ comes from 
consideration of anisotropies in this model and also from the condition for not having the BKL instability. It can be shown that 
the required amount of magnetic field can be generated for value of $p<1$ whereas to avoid BKL instability $p$ should not be greater 
than $1/3$. The scale factor in the equation \eqref{eqn:3.6}, when applied to the equation of motion \eqref{eqn:2.13} along with 
the proper coupling function, leads to the following differential equation to be obeyed by the gauge field 
mode functions.
\begin{equation}\label{eqn:3.7}
\mathcal{A}_{k}''+\Big(k^2-\frac{\gamma(\gamma-1)}{(\tau-\tilde{\tau}_{B-})^2}\Big)\mathcal{A}_k=0
\end{equation}
where $\gamma=\frac{np}{1-p}$. The general solution to the above equation is then given by
\begin{equation}\label{eqn:3.8} 
\mathcal{A}_k(\tau)=\sqrt{-(\tau -\tilde{\tau}_{B-})}\Big[B_1J_{\gamma-\frac{1}{2}}(-k(\tau -\tilde{\tau}_{B-})+B_2Y_{\gamma-\frac{1}{2}}(-k(\tau 
-\tilde{\tau}_{B-})) \Big]
\end{equation}

Here we have chosen Bessel functions of first and second kind as the solution of \eqref{eqn:3.7} for computational convenience.
For the smooth evolution of the mode we need to match the gauge fields at the end of matter dominated phase 
with that of the beginning of Ekpyrotic phase. The junction conditions \cite{Fujita:2016qab} are as follows :
\begin{align}
&\frac{\mathcal{A}_k(\tau)}{I}\bigg |_{M.D.=\tau_E}=\frac{\mathcal{A}_k(\tau)}{I}\bigg |_{Ek=\tau_i}\label{eqn:3.9} \\
&\Big [\frac{\mathcal{A}_k(\tau)}{I}\Big ]'\bigg |_{M.D.=\tau_E}=\Big [\frac{\mathcal{A}_k(\tau)}{I}\Big ]'\bigg |_{Ek=\tau_i}\label{eqn:3.10}
\end{align}
where we have used the abbreviations $M.D.$ and $Ek$ to mean that we have used expressions of the gauge fields of matter 
dominated and Ekpyrotic phases respectively. Here $\tau_E$ is the conformal time when matter dominated phase ends and 
$\tau_i$ is the conformal time when Ekpyrotic phase begins. One must be careful to note that the conformal time is not continuous 
but cosmic time is. Now let us explicitly 
compute \eqref{eqn:3.9},
\begin{multline}\label{eqn:3.11}
 \frac{\sqrt{-\frac{\pi}{4}(\tau-\tilde{\tau}_{E}})H^{(1)}_{\frac{4n-1}{2}}(-k(\tau-\tilde{\tau}_E))}{I}\Bigg |_{\tau_E} \\ 
 =\frac{\sqrt{-(\tau-\tilde{\tau}_{B-})}[B_1J_{\gamma-\frac{1}{2}}(-k(\tau-\tilde{\tau}_{B-})+B_2Y_{\gamma-\frac{1}{2}}(-k(\tau-\tilde{\tau}_{B-})]}{I} 
 \Bigg|_{\tau_i}
\end{multline}
The coupling function should satisfy the equality condition, 
$I \bigg |_{\tau_E}=I \bigg |_{\tau_i}$, because the scale factor is continuous at $\tau_E$ and $\tau_i$ 
as it follows from the condition $a_E \simeq a_{B-} \Big (\frac{H_{B-}}{H_E}\Big )^p$ \cite{Cai:2013vm}.
Thus one obtains from the above equation (\ref{eqn:3.11}),
\begin{multline}\label{eqn:3.12}
\sqrt{-\frac{\pi}{4}(\tau_E-\tilde{\tau}_{E}})H^{(1)}_{\frac{4n-1}{2}}(-k(\tau_E-\tilde{\tau}_E))\\
=\sqrt{-(\tau_i-\tilde{\tau}_{B-})}[B_1J_{\gamma-\frac{1}{2}}(-k(\tau_i-\tilde{\tau}_{B-})+
B_2Y_{\gamma-\frac{1}{2}}(-k(\tau_i-\tilde{\tau}_{B-})]
\end{multline}
Before explicitly calculating \eqref{eqn:3.10} let us  state a specific identity for the Bessel functions which will be useful for the successive 
calculations.
\begin{equation}\label{eqn:3.13}
\frac{f(z)}{f'(z)}J'_\nu(f(z))-\nu J_\nu (f(z))=-f(z)J_{\nu +1}(f(z))
\end{equation}
The above identity is derived using chain rule of differentiation on the identity $J'_\nu(x)-\frac{\nu}{x}J_\nu(x)=-J_{\nu +1}(x)$.
To simplify the above equation \eqref{eqn:3.10} we use the identity \eqref{eqn:3.13} and achieve the following relation.
\begin{multline}\label{eqn:3.14}
\sqrt{-\frac{\pi}{4}(\tau_E-\tilde{\tau}_{E})}H^{(1)}_{\frac{4n+1}{2}}(-k(\tau_E-\tilde{\tau}_E))\\
=\sqrt{-(\tau_i
-\tilde{\tau}_{B-})}\Big[kB_1J_{\gamma+\frac{1}{2}}(-k(\tau_i-\tilde{\tau_{B-}}))+kB_2Y_{\gamma+\frac{1}{2}}(-k(\tau_i-\tilde{\tau}_{B-}))\Big]
\end{multline}
The coefficients  $B_1$ and $B_2$ are obtained by solving \eqref{eqn:3.11} and \eqref{eqn:3.14}. 
\begin{align}
B_1&=\frac{1}{4}k\pi^{\frac{3}{2}}\sqrt{(\tau_i-\tilde{\tau}_{B-})(\tau_E-\tilde{\tau_E})}[Y_{\gamma+\frac{1}{2}}(-k(\tau_i-\tau_{B-}))
H^{(1)}_{2n-\frac{1}{2}}(-k(\tau_E-\tilde{\tau_E}))\nonumber\\
&-Y_{\gamma-\frac{1}{2}}(-k(\tau_i-\tau_{B-}))H^{(1)}_{2n+\frac{1}{2}}(-k(\tau_E-\tilde{\tau_E}))]\label{eqn:3.15}\\
B_2&=\frac{1}{4}k\pi^{\frac{3}{2}}\sqrt{(\tau_i-\tilde{\tau}_{B-})(\tau_E-\tilde{\tau_E})}[-J_{\gamma+\frac{1}{2}}(-k(\tau_i-\tau_{B-}))
H^{(1)}_{2n-\frac{1}{2}}(-k(\tau_E-\tilde{\tau_E}))\nonumber\\
&+J_{\gamma-\frac{1}{2}}(-k(\tau_i-\tau_{B-}))H^{(1)}_{2n+\frac{1}{2}}(-k(\tau_E-\tilde{\tau_E}))]\label{eqn:3.16}
\end{align}
The above expressions for $B_1$ and $B_2$ will be used to compute the power spectrum of the magnetic and electric fields 
in the next section. Studying the electromagnetic power spectrum we will learn about the magnetic fields.
\section{The power spectrum of magnetic and electric fields}
The general form of the power spectrum of electric and magnetic fields respectively are given by
\begin{align}
\mathcal{P}_E(k, \tau) =\frac{d\rho_E}{d\ln k}=\frac{I^2}{2\pi^2} \frac{k^3}{a^4}\Big |\Big (\frac{\mathcal{A}(k,\tau)}{I}\Big )'\Big |^2 \label{eqn:4.1}\\
\mathcal{P}_B(k, \tau) =\frac{d\rho_B}{d\ln k}= \frac{1}{2\pi^2}\Big (\frac{k}{a}\Big )^4k|\mathcal{A}(k,\tau)|^2 \label{eqn:4.2}
\end{align}
The modes which are superhorizon at the end of the Ekpyrotic phase will leave the horizon. They are considered to re enter at present
epoch giving rise to the observable large scale magnetic fields. Now we take the super horizon limit for long wavelengths and use the 
expressions for the gauge fields to obtain the following expressions for the power spectrum.
\begin{align}
\mathcal{P}_E (k, \tau)&=\frac{1}{2\pi^2}\Big(\frac{k}{a}\Big)^4k(\tilde{\tau}_{B-}-\tau)\Big|B_2Y_{\gamma+\frac{1}{2}}(-k(\tau-\tilde{\tau}_{B-}))
\Big|^2\label{eqn:4.3}\\
\mathcal{P_B}(k, \tau)&=\frac{1}{2\pi^2}\Big(\frac{k}{a}\Big)^4k(\tilde{\tau}_{B-}-\tau)|B_2Y_{\gamma -\frac{1}{2}}(-k(\tau-\tilde{\tau}_{B-}))|^2
\label{eqn:4.4}
\end{align}
A careful investigation reveals that the terms $B_1J_{\gamma+\frac{1}{2}}(-k(\tau-\tilde{\tau}_{B-}))$ and 
$B_1J_{\gamma -\frac{1}{2}}(-k(\tau-\tilde{\tau}_{B-}))$ will be less dominant compared to the other parts of the 
solutions when superhorizon modes are being considered. As a result one can  neglect their contribution.
Using the expressions for the gauge field and $B_2$, and expanding for small arguments, we get the expressions for electric and 
magnetic power spectrum. Now we will use the following expressions to simplify \eqref{eqn:4.3} and \eqref{eqn:4.4},
\begin{align}
\tau_i-\tilde{\tau}_{B-}&=\frac{p}{(1-p)}\frac{1}{a_EH_E}\label{eqn:4.5}\\
\tau_E-\tilde{\tau_E}&=\frac{2}{a_EH_E}\label{eqn:4.6}\\
\tau-\tilde{\tau}_{B-}&=\Big(\frac{a}{a_{B-}}\Big)^{\frac{1-p}{p}}\frac{p}{(1-p)}\Big(\frac{1}{a_{B-}H_{B-}}\Big)\label{eqn:4.7}
\end{align}
Thus the spectra of electric and magnetic fields reduce to 
\begin{align}
\mathcal{P}_E&=\frac{1}{4\pi^3}\Big(\frac{-k}{a_{B-}H_{B-}}\Big)^{4-4n}\Big(\frac{a}{a_{B-}}\Big)^{-2n}\exp N(-4n+2\gamma)
\Big(\frac{a_{B-}H_{B-}}{a}\Big)^4\Gamma^2(2n+\frac{1}{2})\label{eqn:4.8}  \\
\vspace{8mm}
\mathcal{P_B}&=\frac{1}{16\pi^3}\Big(\frac{-k}{a_{B-}H_{B-}}\Big)^{6-4n}\exp N(-4n+2\gamma)\Big(\frac{p}{1-p}\Big)^2 \hspace{2mm}
\frac{\Gamma^2(\gamma-\frac{1}{2})\Gamma^2(2n+\frac{1}{2})}{\Gamma^2(\gamma 
+\frac{1}{2})}\Big(\frac{a}{a_{B-}}\Big)^{\frac{2(1-p)}{p}-2n}
\nonumber \\
&\hspace{6mm}\times \Big(\frac{a_{B-}H_{B-}}{a}\Big)^4\label{eqn:4.9}
\end{align}
$H_{B-}$ and $H_E$ are the Hubble parameters in the Ekpyrotic and matter dominated phases respectively. N is defined as number of relative e-foldings,
\begin{equation}\label{eqn:4.10}
N=\ln \Big(\frac{a_{B-}H_{B-}}{a_EH_E}\Big)
\end{equation}
Considering the constraints on $n$ to deal with strong coupling problem and that on $p$ from anisotropy study and instability issues, 
a detailed analysis of the terms involved we choos $\gamma=\frac{np}{1-p}> \frac{1}{2}$.  We will soon show that the scale 
invariance of the magnetic field spectrum occurs at $n=\frac{3}{2}$ and $p$ is already constrained by CMB anisotropies and 
BKL instability.

\section{Predicting magnetic fields of observable strength}
In the previous section we have shown elaborately how the scale invariant magnetic field spectrum can be obtained in non singular bouncing 
cosmology. Now we use equation \eqref{eqn:4.9} to predict the present day magnetic field $B_0$. We first find the 
magnetic field power spectrum at the end of Ekpyrotic era.  Based on the argument given earlier we assume the effect of 
the fast bounce to be limited and hence we treat this phase as instantaneous. As the coupling function $I$ reduces to unity 
at the end of Ekpyrotic phase the vector field behaves like a Maxwellian field and evolves as the square of the scale factor. 
One may thus consider that effectively the radiation dominated era start from $\tau_{B-}$. Thus the magnetic field strength 
observed today is given by 
\begin{align}\label{eqn:5.1}
B_0=&(\mathcal{P}_B(k,a_0))^\frac{1}{2}=\frac{1}{4\pi^{\frac{3}{2}}}
\exp N\Big(-3+\frac{3p}{2(1-p)}\Big)\Big (\frac{p}{1-p}\Big )
\frac{\Gamma\Big(\frac{3p}{2(1-p)}-\frac{1}{2}\Big)\Gamma(\frac{7}{2})}{\Gamma\Big(\frac{3p}{2(1-p)}+
\frac{1}{2}\Big)} \Big (\frac{a_0}{a_{B-}}\Big )^{\frac{1-p}{p}-\frac{3}{2}}\nonumber \\
&\times \Big (\frac{a_{B-}H_{B-}}{a_0}\Big )^2
\end{align}  
Following the conservation of entropy one can relate the power spectrum at the end of the Ekpyrotic phase to that of observed 
today and the scale factors of the corresponding epochs are related via the temperatures.
\begin{equation}\label{eqn:5.2}
\frac{a_{B-}}{a_0}=\frac{T_0}{\rho_{em}(a_{B-})^{\frac{1}{4}}}=\frac{T_0}{T_{B-}}
\end{equation}
where $a_0$ is the scale factor today and $T_0$ the temperature. At the end of Ekpyrotic phase the temperature is 
$T_{B-}$ and the electromagnetic energy density is given by $\rho_{em}$.
For super horizon scales, it is found that the density of the electric field dominates over that of the magnetic field.
Hence we will only consider the contribution of electric field while calculating the electromagnetic energy density. Thus 
\begin{equation}\label{eqn:5.3}
\rho_{em}=\int_{a_EH_E}^{a_{B-}H_{B-}}\frac{dk}{k}\frac{1}{4\pi^3}\Big(\frac{-k}{a_{B-}H_{B-}}\Big)^{4-4n}\Big(\frac{a}{a_{B-}}
\Big)^{-2n}\exp N
(-4n+2\gamma)\Big(\frac{a_{B-}H_{B-}}{a}\Big)^4\Gamma^2(2n+\frac{1}{2})
\end{equation}

We are interested in the scale invariant case at the end of Ekpyrotic phase and hence \eqref{eqn:5.3} reduces to,
\begin{equation}\label{eqn:5.4}
\rho_{em}(a_{B-})=\frac{1}{8\pi^3}\exp\Big(-4N+\frac{3p}{(1-p)}\Big)(H_{B-})^4\Gamma^2\Big(\frac{7}{2}\Big)
\end{equation}

Here we have made the approximation $a_EH_E\ll a_{B-}H_{B-}$.
Now putting these in entropy conservation equation \eqref{eqn:5.2} we get,
\begin{align}
\Big(\frac{a_{B-}}{a_0}\Big)^2 &=\frac{T_0^2}{\rho_{em}(a_{B-})^{\frac{1}{2}}}\nonumber \\
&=\frac{T_0^2 2\sqrt{2}\pi^{\frac{3}{2}}\exp(2N-\frac{3Np}{2(1-p)})}{H_{B-}^2\Gamma(\frac{7}{2})}\label{eqn:5.5}\\
\Big(\frac{a_{B-}H_{B-}}{a_0}\Big)^2&=\frac{T_0^2 2\sqrt{2}\pi^{\frac{3}{2}}\exp(2N-\frac{3Np}{2(1-p)})}{\Gamma(\frac{7}{2})}\label{eqn:5.6}
\end{align}
Therefore the present day magnetic field can be calculated using the above expression in equation \eqref{eqn:5.1},
\begin{equation}\label{eqn:5.7}
B_0=\frac{1}{4\pi^{\frac{3}{2}}}\Big(\frac{p}{1-p}\Big)\exp(-N)\frac{\Gamma\Big(\frac{3p}{2(1-p)}-\frac{1}{2}\Big)\Gamma(\frac{7}{2})}
{\Gamma\Big(\frac{3p}{2(1-p)}+\frac{1}{2}\Big)}\Big(\frac{a_0}{a_{B-}}\Big)^{\frac{1-p}{p}-\frac{3}{2}}\hspace{4mm}\frac{T_0^2 2\sqrt{2}
\pi^{\frac{3}{2}}}{\Gamma(\frac{7}{2})}
\end{equation}
Note that $N$ is the relative e-foldings as defined in eq. \eqref{eqn:4.10} and values of the parameter $p$ of the model will 
be chosen in such a way that it satisfies the bound coming from the analysis of anisotropies in this model and also takes care of the 
BKL instability. From the first the requirement is $p < 1$ and we have found that one achieve observationally interesting magnetic field
within that limit. But the BKL instability will appear for values of $p$ larger than 1/3. The temperature of the universe today is $2.73$K 
so putting this in \eqref{eqn:5.7} and simplifying we get,
\begin{equation}\label{eqn:5.11}
B_0=\frac{1}{\sqrt{2}}\Big(\frac{p}{1-p}\Big)\exp(-N)\frac{\Gamma\Big(\frac{3p}{2(1-p)}-\frac{1}{2}\Big)}{\Gamma\Big(\frac{3p}{2(1-p)}
+\frac{1}{2}\Big)}\Big(\frac{a_0}{a_{B-}}\Big)^{\frac{1-p}{p}-\frac{3}{2}}2.838 \times 10^{-6}~\text{Gauss}
\end{equation}
One of the major issues in bouncing cosmology is that the energy scales are not well defined. In the following Figure \ref{Fig 2.} we 
plot the result obtained in Eq. \eqref{eqn:5.11} by assuming a value of $N=50$ and see that our model can generate magnetic fields 
of observational interest (greater the lower bound of $B_0 \ge 3\times 10^{-16}$ Gauss). 
\begin{figure}[h]
\centering
\includegraphics[width = 8 cm, height = 4.5 cm]{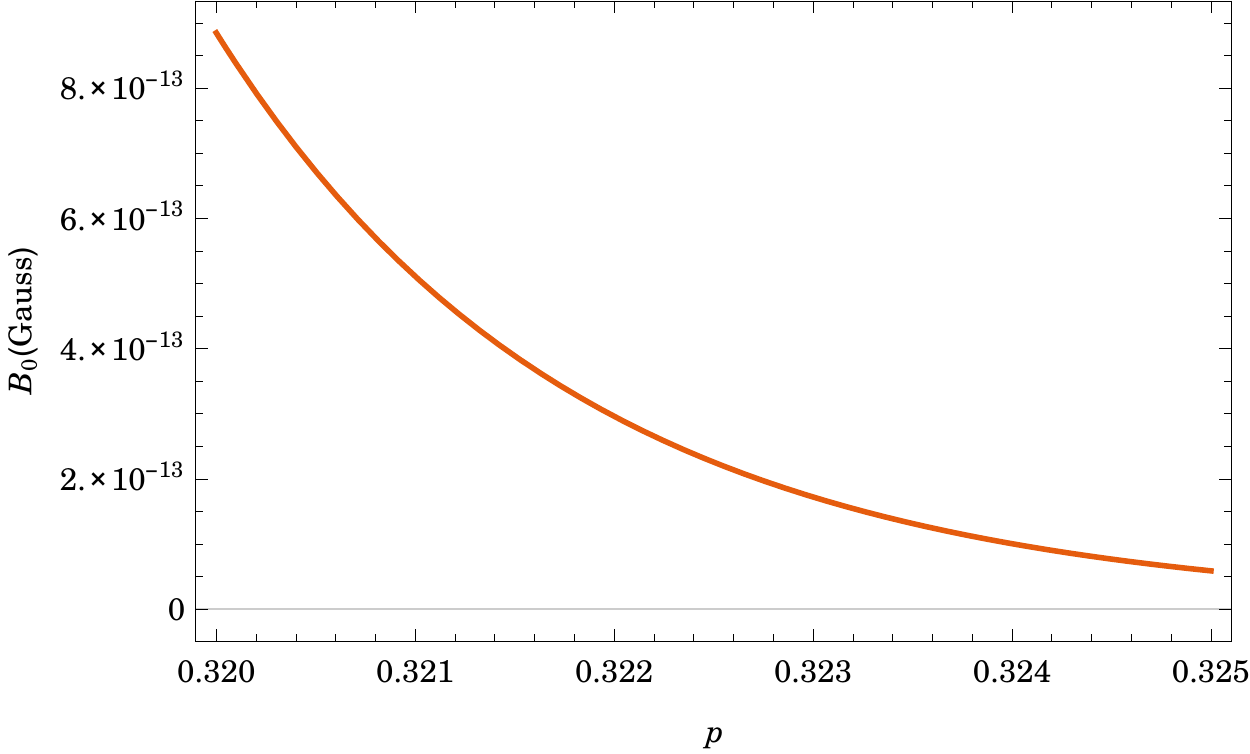}
\hfill
\caption{\label{Fig 2.} The plot shows that magnetic field strength of observational interest can be achieved in the model under 
concern with a choice for effective number of efoldings $N=50$ and within the allowed range of $p$.}
\end{figure}
To be compatible with the anisotropies observed in CMB by Planck satellite what we need is a very small contribution to the 
density. This in turn puts a bound \cite{Cai:2013vm}  on the scale of Ekpyrotic phase. Though it is not well settled like the minimal inflationary scenario, 
still we use this as a guide line for fixing the value of $N$. Let us write the expression for the number of e-foldings $N$ by considering
the above bound as, 
\begin{equation}\label{eqn:5.12}
N>\frac{1-p}{2(1-3p)}\ln\Big(\frac{1}{\Omega_{pert}} \Big) 
\end{equation}
where the amount of energy density in matter perturbations, $\Omega_{pert} \sim \mathcal{O}(10^{-5})$. Now putting this bound 
in eq. \eqref{eqn:5.11} we get,
\begin{equation}\label{eqn:5.13}
B_0<\frac{1}{\sqrt{2}}\Big(\frac{p}{1-p}\Big)(\Omega_{pert})^{\frac{1-p}{2(1-3p)}}\frac{\Gamma\Big(\frac{3p}{2(1-p)}-\frac{1}{2}\Big)}
{\Gamma\Big(\frac{3p}{2(1-p)}+\frac{1}{2}\Big)}\Big(\frac{a_0}{a_{B-}}\Big)^{\frac{1-p}{p}-\frac{3}{2}}  2.838 \times 
10^{-6}~\text{Gauss}
\end{equation}
Let us now determine $\Big(\frac{a_0}{a_{B-}}\Big)$. We know that the scale factors are inversely proportional to the temperature, hence 
$\frac{a_0}{a_{B-}}=\frac{T_{B-}}{T_0}$. Taking present temperature of the Universe, $T_0=10^{-4}$ eV and the Ekpyrotic 
temperature, $T_{B-}=10^{20}$ eV following \cite{Khoury:2001wf} the inequality in eq. \eqref{eqn:5.13} becomes,
\begin{equation}\label{eqn:5.14}
B_0<\frac{1}{\sqrt{2}}\Big(\frac{p}{1-p}\Big)10^{-\frac{5(1-p)}{2(1-3p)}}\frac{\Gamma\Big(\frac{3p}{2(1-p)}-\frac{1}{2}\Big)}
{\Gamma\Big(\frac{3p}{2(1-p)}+\frac{1}{2}\Big)}10^{\frac{24(1-p)}{p}-36}\hspace*{2mm}2.838\times 10^{-6}~\text{Gauss}
\end{equation}
The model parameter $p$ gets further constrained if the observational bound on primordial magnetic field from Planck 2015 
is imposed in the above expression. In figure \ref{Fig 1.} we have shown  that the observational constraint for nearly scale invariant 
magnetic field without considering the interaction with matter fields - $B_0<2\times 10^{-9}$  Gauss \cite{Ade:2015cva} 
is satisfied within the range $ 0.3067 < p < 0.3077$.
\begin{figure}[h]
\centering
\includegraphics[width = 8 cm, height = 4.5 cm]{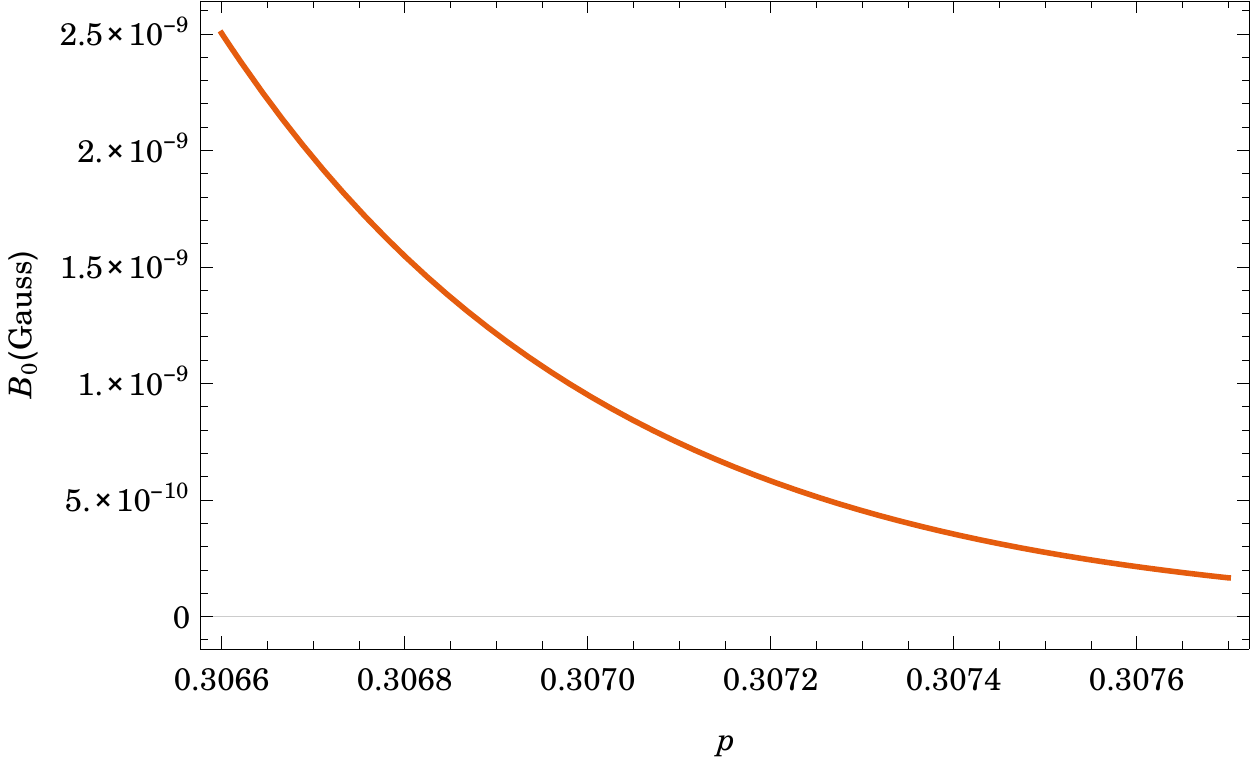}
\hfill
\caption{\label{Fig 2.} The plot shows the variation of magnetic field $B_0$ with the parameter $p$ and depicts the allowed range of the same 
constrained by Planck 2015 data.}
\end{figure}
Thus we obtain the magnetic field strength effective foe recent blazar observations in the scenario of a non singular 
bouncing cosmology. In our analysis so far, we have assumed that the energy density of the produced electromagnetic field 
is not large enough to alter the background evolution.  In the following section we study the back reaction problem in details and 
impose this condition to the obtained result.

\section{Addressing the back reaction problem}
It is important to check that the electromagnetic energy produced during magnetogenesis do not spoil the background dynamics. 
In other words, we need to check that the electromagnetic energy density is less than the energy density responsible for 
the background evolution. As we have discussed in the previous section, the produced magnetic field density is much less compared
to the electric field density. So for current consideration it will suffice to consider only the electric field. The expression for 
electromagnetic field energy density $\rho_{em}(a_{B-})$ has been shown in \eqref{eqn:5.4}. The condition to avoid the 
back reaction problem one needs to satisfy the following condition.
\begin{align}
\rho_{em}(a_{B-})&<3H^2_{B-}M^2_{pl}\nonumber \\
\frac{1}{8\pi^3}\exp\left(-4N+\frac{3p}{1-p}\right)(H_{B-})^4\Gamma^2\left(\frac{7}{2}\right)&<3H^2_{B-}M^2_{pl}\label{eqn:6.1} \\
\mid H_{B-}\mid & <  \frac{2\sqrt{6}\pi^{\frac{3}{2}}\exp\left(2N-\frac{3p}{2(1-p)}\right)}{\Gamma(\frac{7}{2})}M_{pl} \label{eqn:6.2}
\end{align}
We see that for a sucessful magnetogenesis the above inequality must have to be satisfied. 
As there is no definite prediction for $H_{B-}$ for the model under concern we can thus put an upper bound on the energy scale 
following the above inequality and demand that sucessful magnetogenesis without the back reaction and strong coupling problem 
can be achieved if such a scale exists. However independent cross check is utterly necessary to establish such a claim in a firm footing.

\section{Conclusion}
Let us conclude by summerising the results obtained by us. In this work we have investigated primordial magnetogenesis in a model 
of non singular bouncing cosmology which has an Ekpyrotic phase preceded by an initial matter contraction phase. The bounce and 
fast roll have been considered to be instantaneous. The background evolution of this model is governed by a pressure less matter fluid 
in the early matter contraction phase whereas the evolution of the Ekpyrotic phase is governed by a scalar field with a non standard 
kinetic term and a non trivial potential. As most of the bouncing models suffer from production of large amount of anisotropies giving rise 
to huge non Gaussianities they become non viable from the present observational view point.  However the model we have considered here 
has two steps of contraction phase and as an advantage of the additional phase the extra anisotropies die down making the 
model viable in the current observational scenario after Planck 2015. 

For the generation of primordial magnetic fields the necessary conformal symmetry breaking of the electromagnetic Lagrangian has been achieved via a coupling of the kinetic term with a 
non background scalar field. To achieve the desired amount of magnetic field compatible with present observational data the model parameters 
are tuned by constraints coming from study of anisotropy in matter fluctuations. The BKL instability have also been taken care of while choosing 
those parameters. A general picture of magnetogenesis has been represented here by considering the electromagnetic 
mode functions being generated in the matter dominated phase and rolled down through the Ekpyrotic phase. With proper 
choice of boundary conditions we have calculated the power spectrum for both the electric and magnetic fields. Remarkably a scale 
invariant power spectrum for the magnetic field has been achieved within the above mentioned parameter range. 
One of the  major issues with the bouncing cosmology is that the scales of each evolution phase are not well settled. In this work 
we have imposed an upper bound on the scale of Ekpyrotic phase which is responsible for magnetogenesis without the back reaction problem. 
The other important scale is the duration of contraction phase, for this we have used the prescription which supports the solution of monopole 
and flatness problem in this bouncing model. Though not well specified, the parameters and scales those have been used here show a harmony 
in satisfying several observational constraints which is remarkable. 

The results obtained in  this work are quite satisfactory. However there still remain several issues that one should address to have a complete 
picture. For example, the produced field inevitably contributes to the curvature perturbation. A detailed study is required since the 
electromagnetic spectra is scale dependent in almost all the phases during the evolution and it is important to check that it remains sub dominant
compared to the scale invariant vacuum fluctuations to remain compatible with CMB observations. Moreover we have considered an instantaneous
phase of bounce. It remains very important to check what happens to the modes if we relax the above condition and allow the modes to pass through 
the bounce before exiting the horizon.  This we leave for a future study. Also, inflation being the most accepted era of early universe cosmology it is really challenging for other 
alternatives to predict new observational effects that will help one to distinguish between models of early universe cosmology. We would 
also like to study if the primordial magnetogenesis is really a phenomenon in bouncing cosmology. It will be interesting to study 
the specific signature that it leaves which can be probed in future observations.

\section{Acknowledgement}
We would like to thank Yi-Fu Cai for useful comments and suggestions. SS thanks Presidency University, Kolkata for a visit during 
which this work was done. 

\bibliography{draftjv}
\bibliographystyle{JHEP}

\end{document}